\makeatletter
\input{filecontents.sty}
\makeatother

\documentclass[%
titlepage,
charprotruding,
secnumarabic,
showpacs,
twocolumn,
titlepageabstract,
twoside,aps,pre]{oxprepr}

\def\Preprint{}

\newenvironment{acknowledgement}{\acknowledgments}{}
\makeatletter
\appdef\tabular@hook{\def\@halignto{to\hsize}}%
\let\tableft@skip@default\tableft@skip
\let\tableft@skip\tableft@skip@float
\let\tabmid@skip@default\tabmid@skip
\let\tabmid@skip\tabmid@skip@float
\let\tabright@skip@default\tabright@skip
\let\tabright@skip\tabright@skip@float

\let\orig@abstract\abstract
\let\orig@endabstract\endabstract
\renewcommand{\abstract}[1]{\orig@abstract#1\orig@endabstract}

\let\orig@email\email%
\def\email#1{\gdef\my@email{#1}}%
\newcommand{\institute}[1]{%
  \let\newline\null%
  \affiliation{#1}%
  \orig@email[E-mail:]{\my@email}%
}

\makeatother

\makeatletter
\ifx\@undefined\Preprint
  \def\JournalPreprint#1#2{#1}
\else
  \def\JournalPreprint#1#2{#2}
\fi
\makeatother

\JournalPreprint{\documentclass[epj]{svjour}}{}

\usepackage[T1]{fontenc}
\usepackage[american]{babel}
\usepackage{amsfonts}
\usepackage{amsmath}
\usepackage{amssymb}
\usepackage{graphicx}
\usepackage[pdftitle={Parameter estimation for power-law distributions by maximum likelihood methods},
pdfauthor={Heiko Bauke},
pdfcreator={},
pdfsubject={},
pdfproducer={},
pdfkeywords={power-law},
hyperfootnotes=false,naturalnames]{hyperref}
\usepackage{url}
\usepackage[Symbol]{upgreek}

\graphicspath{{figures/}}

\InputIfFileExists{date.tex}{}

\newcommand*{\e}{\mathrm{e}}
\newcommand*{\D}{\mathrm{d}}
\newcommand*{\N}{\mathbb{N}}
\newcommand*{\Prob}[1]{\mathbb{P}\left[#1\right]}
\newcommand*{\E}[1]{\mathbb{E}\left[#1\right]}

\newcommand*{\Id}[1]{\mathbb{I}\left[#1\right]}
\newcommand{\argmax}{\operatornamewithlimits{argmax}}
\newcommand{\round}{\operatorname{round}}
\JournalPreprint{%
  \newcommand{\definedas}{\mathrel{\raise.095ex\hbox{:}\mkern-4.2mu=}}%
}{%
  \newcommand{\definedas}{\mathrel{\raise.04ex\hbox{:}\mkern-4.2mu=}}%
}
\makeatletter
\newcommand*{\Vdots}{%
  \settowidth{\@tempdimb}{$=$}%
  \mathrel{\makebox[\@tempdimb]{\vdots}}%
}
\makeatother

\begin{document}

\newcommand*{\Figurename}{Figure}
\newcommand*{\Tablename}{Table}

\title{Parameter estimation for power-law distributions\\
  by maximum likelihood methods}

\author{Heiko Bauke}
\institute{Rudolf Peierls Centre for Theoretical Physics, University of
  Oxford, 1~Keble~Road, Oxford, OX1~3NP, United~Kingdom\newline
  \email{heiko.bauke@physics.ox.ac.uk}}
\JournalPreprint{%
}{%
}
\abstract{%
  Distributions following a power-law are an ubiquitous phenomenon. Methods
  for determining the exponent of a power-law tail by graphical means are
  often used in practice but are intrinsically unreliable. Maximum likelihood
  estimators for the exponent are a mathematically sound alternative to
  graphical methods.}%

\date{\today}


\maketitle

\section{Introduction}
\label{sec:intro}

The distribution of a discrete random variable is referred to as a
distribution with a power-law tail if it falls as
\begin{equation}
  \label{eq:tail}
  p(k) \sim k^{-\gamma}
\end{equation}
for $k\in\N$ and $k\ge k_\mathrm{min}$. Power-laws are ubiquitous
distributions that can be found in many systems from different disciplines,
see \cite{Newman:2005:power-laws,Keller:2005:scale-free_networks} and
references therein for some examples.

Experimental data of quantities that follow a power-law are usually very
noisy; and therefore obtaining \emph{reliable} estimates for the exponent
$\gamma$ is notoriously difficult. Estimates that are based on graphical
methods are certainly used most often in practice. But simple graphical
methods are intrinsically unreliable and not able to establish a reliable
estimate of the exponent $\gamma$.

For that reason, the authors of \cite{Goldstein:etal:2004:fitting_power-law}
introduced an alternative approach based on a maximum likelihood estimator for
the exponent $\gamma$.  Unfortunately the authors concentrate on a rather
idealized type of power-law distributions, namely
\begin{equation}
  \label{eq:p1}
  p_1(k; \gamma) = \frac{k^{-\gamma}}{\zeta(\gamma, 1)}
\end{equation}
with $k\in\N$, where the normalization constant $\zeta(\gamma, 1)$ is given by
the Hurwitz-$\zeta$-function which is defined for $\gamma>1$ and $a>0$ by
\begin{equation}
  \label{eq:H-zeta}
  \zeta(\gamma, a) = \sum_{i=0}^\infty \frac{1}{(i+a)^\gamma}\,.
\end{equation}
The distribution (\ref{eq:p1}) is characterized by one parameter only, and
therefore all properties of this distribution (e.\,g. its mean) are determined
solely by the exponent $\gamma$. In many applications the power-law
(\ref{eq:p1}) is too restrictive.

If one states that a quantity follows a power-law, then this means usually
that the \emph{tail} ($k\ge k_\mathrm{min}$) of the distribution $p(k)$ falls
proportionally to $k^{-\gamma}$. Probabilities $p(k)$ for $k<k_\mathrm{min}$
may differ from the power-law and admit the possibility to tune the mean or
other characteristics independently of $\gamma$. In some situations
probabilities $p(k)$ may differ from a power-law for $k\ge k_\mathrm{max}$ as
well, e.\,g.\ the distribution may have an exponential cut-off.

Therefore, I will generalize the maximum likelihood approach introduced in
\cite{Goldstein:etal:2004:fitting_power-law} to distributions that follow a
power-law within a certain range $k_\mathrm{min}\le k < k_\mathrm{max}$ but
differ from a power-law outside this range in an arbitrary way. Furthermore, I
will give some statements about the large sample properties of the estimate of
the power-law exponent and present a numerical procedure to identify the
power-law regime of the distribution $p(k)$. But first, let us see what is
wrong with popular graphical methods.

\section{Trouble with graphical methods}
\label{sec:garphical}

\sloppy All graphical methods for estimating power-law exponents are based on
a linear least squares fit of some empirical data points $(x_1, y_1)$, $(x_2,
y_2)$,\dots, $(x_M, y_M)$ to the function
\begin{equation}
  y(x) = a_0 + a_1 x \,.
\end{equation}
\fussy The linear least squares fit minimizes the residual
\begin{equation}
  \label{eq:residual}
  \Delta = \sum_{i=1}^M (y_i - a_0 - a_1 x_i)^2 \,.
\end{equation}
Estimates $\hat a_0$ and $\hat a_1$ of the parameters $a_0$ and $a_1$ are
given by \cite{Ramsey:Schafer:2002:Sleuth}
\begin{align}
  \label{eq:a_0}
  \hat a_0 & = \frac{ 
    \left(\sum_{i=1}^M y_i\right)
    \left(\sum_{i=1}^M x_i^2\right) - 
    \left(\sum_{i=1}^M x_i\right)
    \left(\sum_{i=1}^M y_i x_i\right) }{ 
    M
    \left(\sum_{i=1}^M x_i^2\right) - 
    \left(\sum_{i=1}^M x_i\right)^2
  }   \displaybreak[1]\\
  \intertext{and}
  \label{eq:a_1}
  \hat a_1 & = \frac{ 
    M
    \left(\sum_{i=1}^M y_i x_i\right) - 
    \left(\sum_{i=1}^M x_i\right)
    \left(\sum_{i=1}^M y_i\right) }{ 
    M
    \left(\sum_{i=1}^M x_i^2\right) - 
    \left(\sum_{i=1}^M x_i\right)^2 }\,.
\end{align}

The ansatz for the residual (\ref{eq:residual}) and derivation of
(\ref{eq:a_0}) and (\ref{eq:a_1}) are based on several assumptions regarding
the data points $(x_i, y_i)$. It is assumed that there are no statistical
uncertainties in $x_i$, but $y_i$ may contain some statistical error. The
errors in different $y_i$ are independent identically distributed random
variables with mean zero. In particular the standard deviation of the error is
independent of $x_i$. For various graphical methods for the estimation of the
exponent of a power-law distribution these conditions are not met, leading to
the poor performance of these methods.

\begin{figure*}
  \centering%
  \begin{minipage}[c]{0.03\linewidth}%
  \end{minipage}%
  \begin{minipage}[c]{0.485\linewidth}
    \centering%
    \hspace*{1.6cm}\makebox[0pt]{fit on the distribution $\hat p(k)$}%
  \end{minipage}%
  \begin{minipage}[c]{0.485\linewidth}
    \centering%
    \hspace*{1.6cm}\makebox[0pt]{fit on the cumulative distribution $\hat P(k)=\sum_{i\ge k}\hat p(i)$}%
  \end{minipage}\\[3ex]%
  \begin{minipage}[c]{0.03\linewidth}%
    \rotatebox{90}{no binning}%
  \end{minipage}%
  \begin{minipage}[c]{0.485\linewidth}
    \centering%
    \raisebox{3ex}{a)} \includegraphics[width=0.9\linewidth]{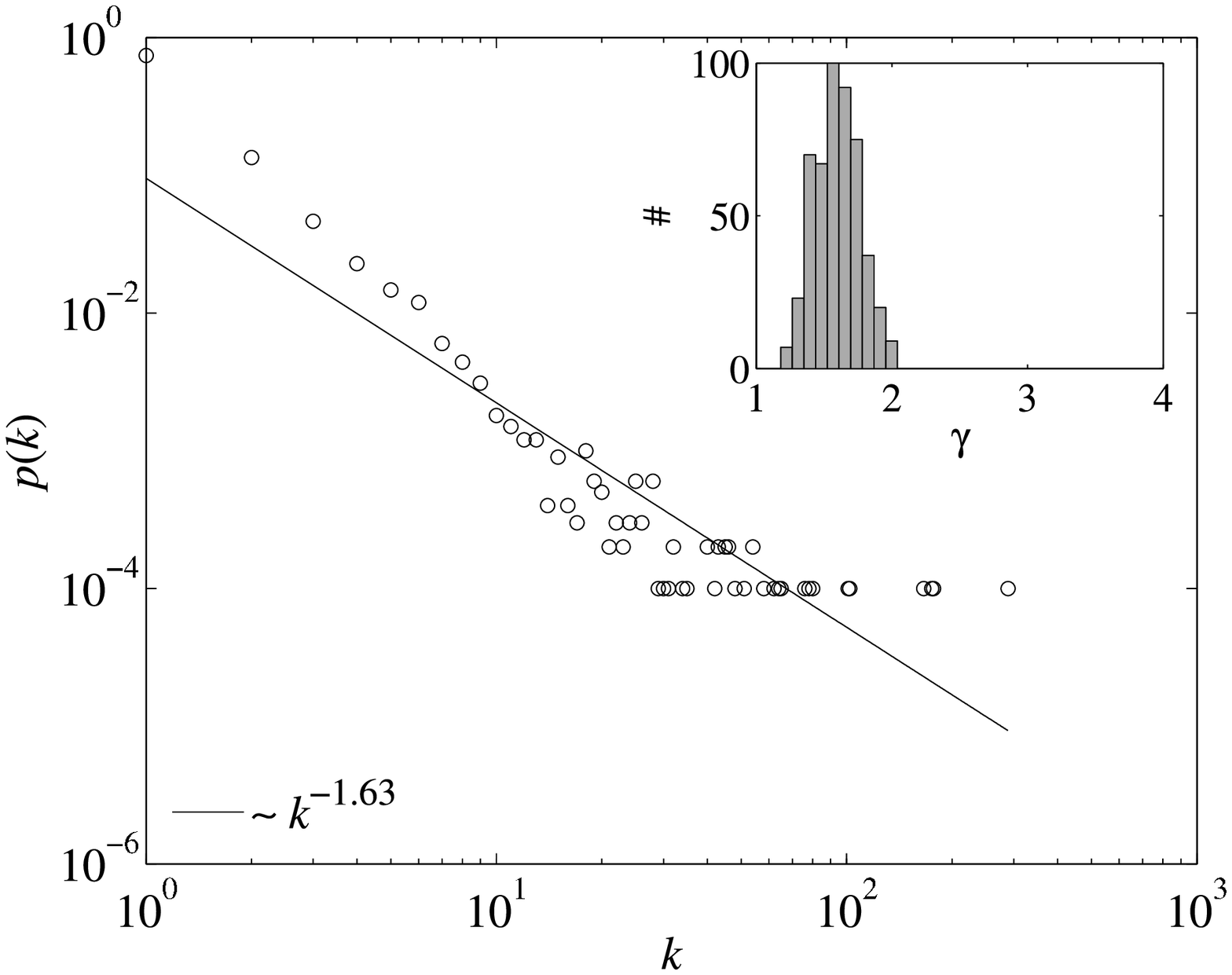}%
  \end{minipage}%
  \begin{minipage}[c]{0.485\linewidth}
    \centering%
    \raisebox{3ex}{b)} \includegraphics[width=0.9\linewidth]{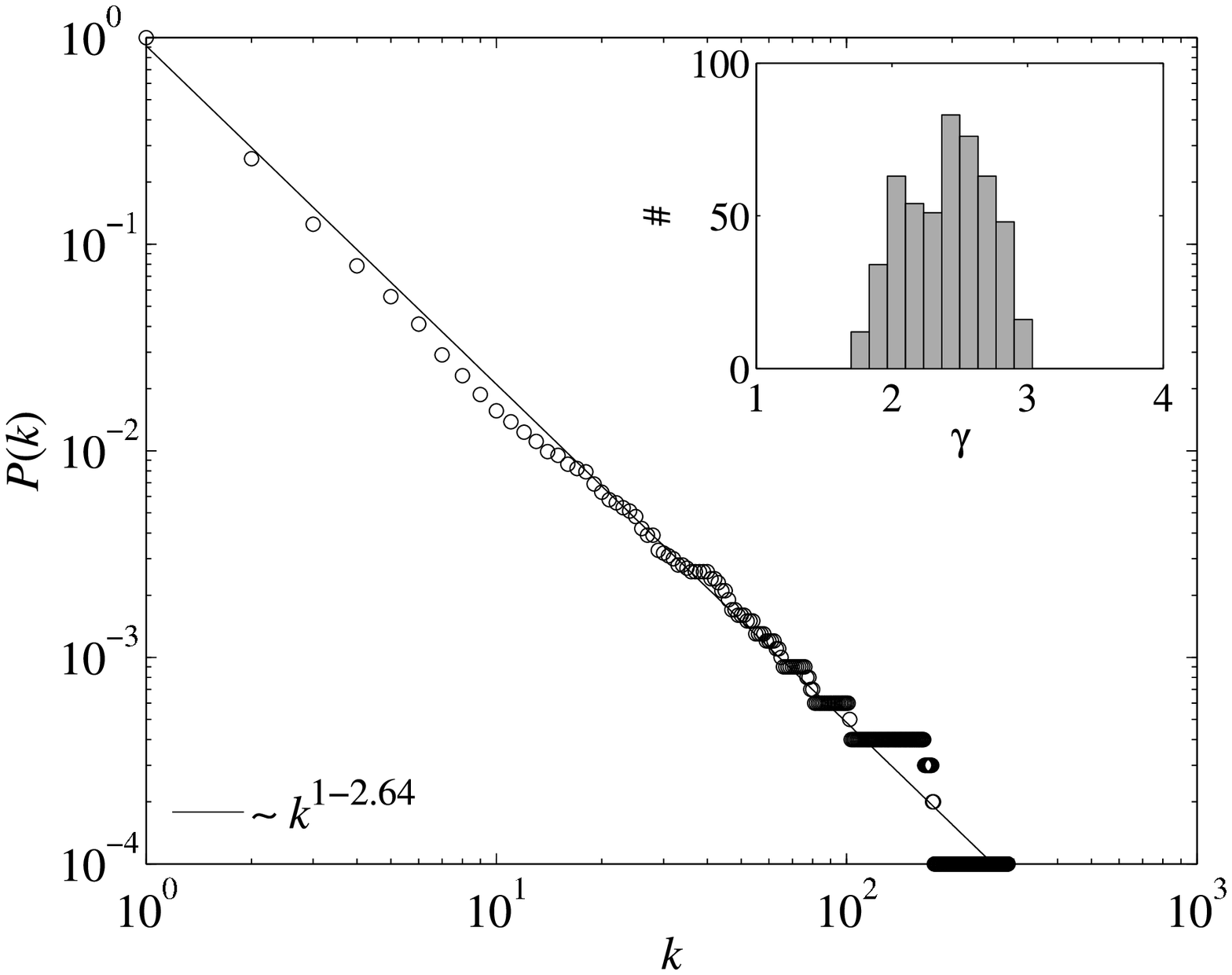}%
  \end{minipage}\\[3ex]%
  \begin{minipage}[c]{0.03\linewidth}%
    \rotatebox{90}{logarithmic binning}%
  \end{minipage}%
  \begin{minipage}[c]{0.485\linewidth}
    \centering%
    \raisebox{3ex}{c)} \includegraphics[width=0.9\linewidth]{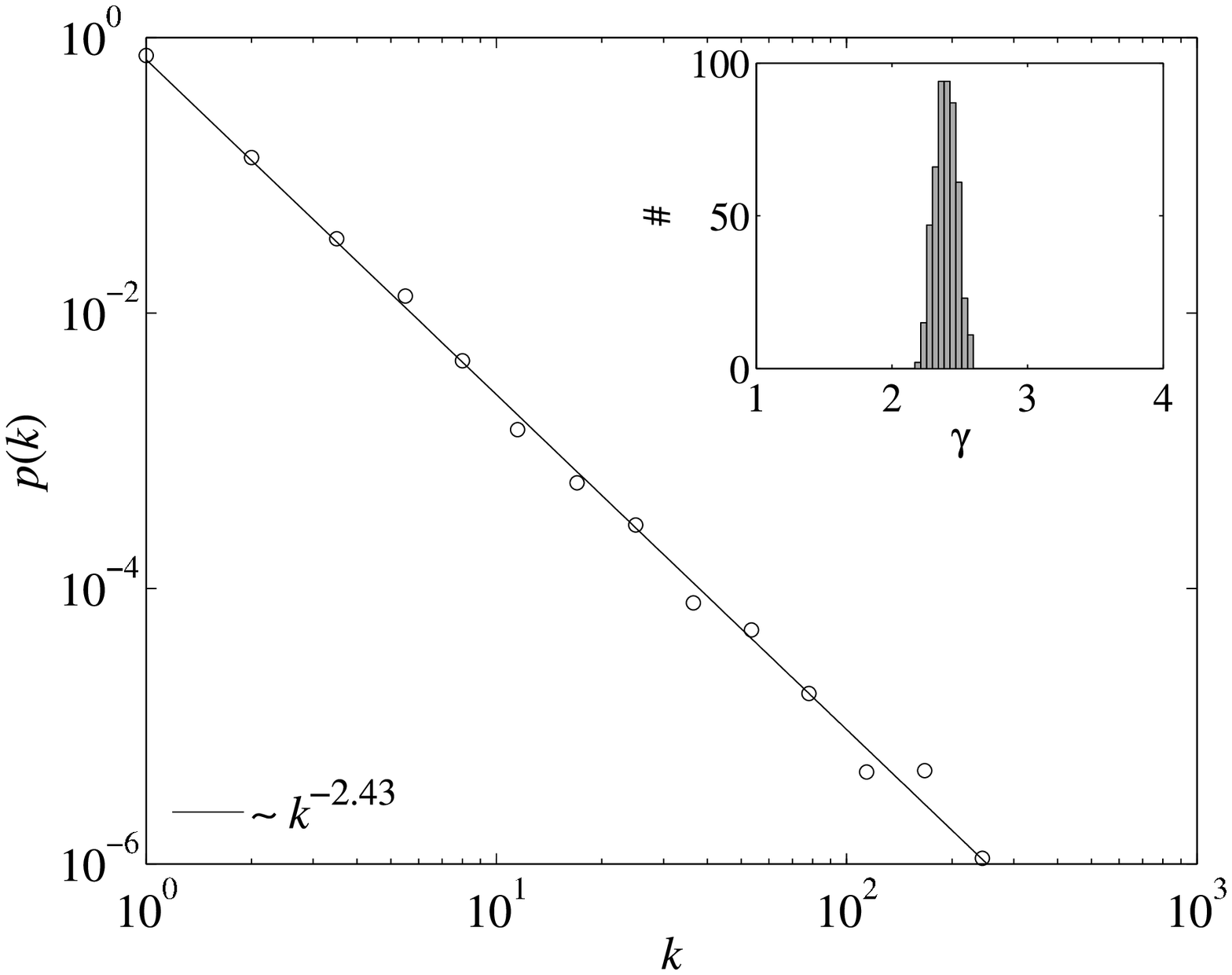}%
  \end{minipage}%
  \begin{minipage}[c]{0.485\linewidth}
    \centering%
    \raisebox{3ex}{d)} \includegraphics[width=0.9\linewidth]{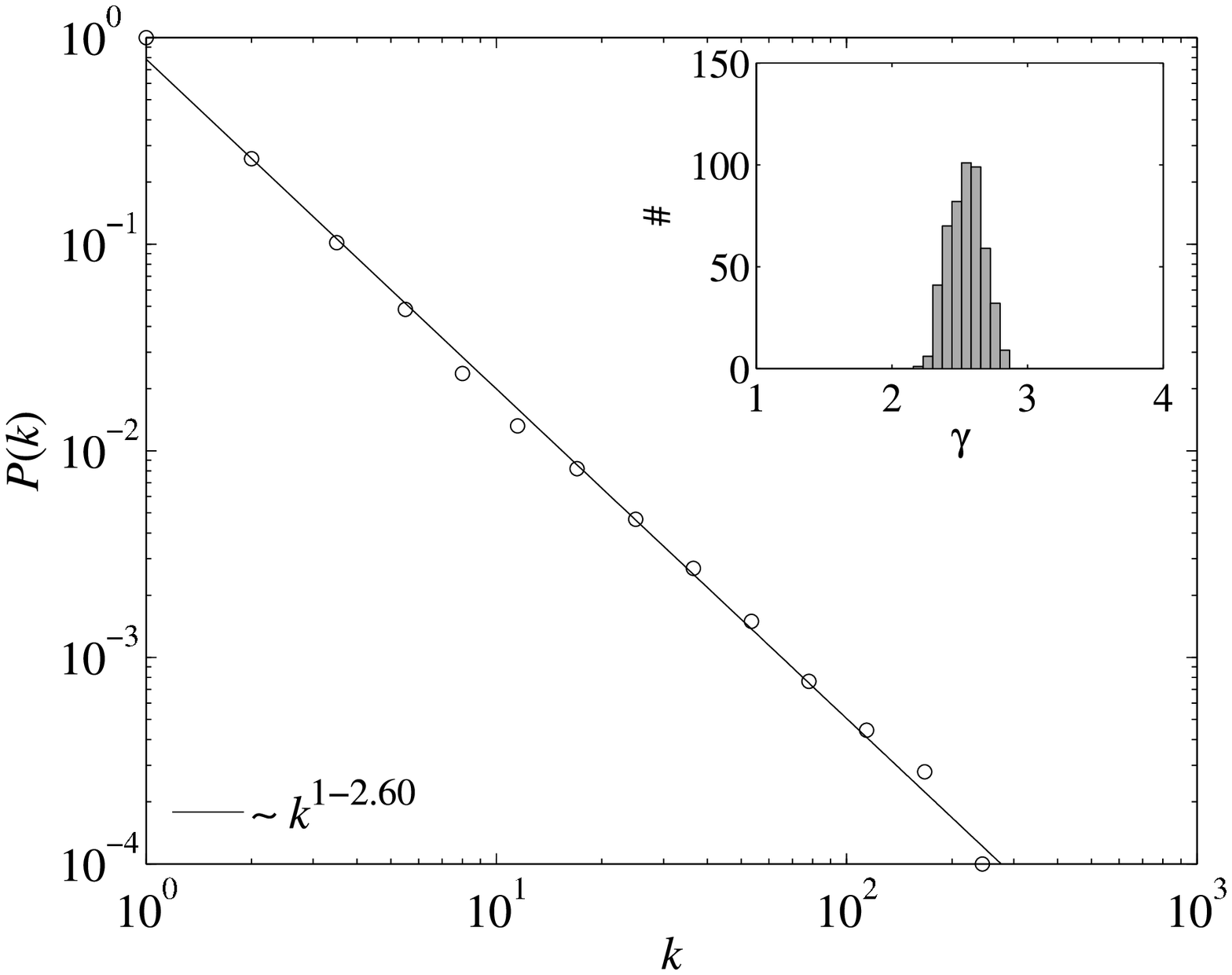}%
  \end{minipage}
  \caption{Comparison of various methods for estimating the exponent of a
    power-law. Each figure shows data for a single data set of $N=10\,000$
    samples drawn from distribution (\ref{eq:p1}) with $\gamma=2.5$. Insets
    present histograms of estimates for $\gamma$ for 500 different data sets.}
  \label{fig:graphical}
\end{figure*}

To illustrate the failure of graphical methods by a computer experiment
$N=10\,000$ random numbers $m_i$ had been drawn from distribution
(\ref{eq:p1}) with $\gamma=2.5$ and an estimate $\hat\gamma$ for the exponent
$\gamma$ was determined by various graphical methods. The estimator is a
random variable and its distribution depends on the method that has been used
to obtain the estimate. Important measures of the quality of an estimator are
its mean and its standard deviation. If the mean of the estimator equals the
true exponent $\gamma$ then the estimator is unbiased and estimators with a
distribution that is concentrated around $\gamma$ are desirable.  For each
graphical method a histogram of the distribution of the estimator was
calculated to rate the quality of the estimator by repeating the numerical
experiment 500 times.

The most straight forward (and most unreliable) gra\-phical approach is based
on a plot of the empirical probability distribution $\hat p(k)$ on a
double-logarithmic scale. Introducing the indicator function $\Id{\cdot}$,
which is one if the statement in the brackets is true and else zero, the
empirical probability distribution is given by
\begin{equation}
  \label{eq:epirical}
  \hat p(k) = \frac{1}{N} \sum_{i=1}^N \Id{m_i=k}\,.
\end{equation}
An estimate $\hat\gamma$ for the power-law exponent $\gamma$ is established by
a least squares fit to
\begin{equation}
  (x_i, y_i) = (\ln k, \ln\hat p(k))
  \quad\text{for all $k\in\N$ with $\hat p(k)>0$,}
\end{equation}
$\hat\gamma$ equals the estimate (\ref{eq:a_1}) for the slope, see
\figurename~\ref{fig:graphical}\,a. Because the lack of data points in the
tail of the empirical distribution this procedure underestimates
systematically the exponent $\gamma$, see
\tablename~\ref{tab:estimate_statistics}.

\begin{table}
  \caption{Mean and standard deviation of the distribution of the estimate for
    the power-law exponent $\gamma$ for various methods. All methods have been
    applied to the same data sets of random numbers from distribution
    (\ref{eq:p1}) with $\gamma=2.5$. See text for details.} 
  \label{tab:estimate_statistics}
  \renewcommand{\thefootnote}{\thempfootnote}%
  \renewcommand{\footnoterule}{}%
  \begin{minipage}{1.0\linewidth}
    \renewcommand{\arraystretch}{1.2}%
    \begin{tabular}{@{}p{30ex}rr@{}}
      \hline\hline
      & mean & standard deviation \\[-0.75ex]
      method & estimate & of estimate \\
      \hline
      fit on empirical distribution & 1.597 & 0.167 \\
      fit on cumulative empirical distribution & 2.395 & 0.304 \\
      fit on empirical distribution with logarithmic
      binning\protect\footnote{In 
        \cite{Goldstein:etal:2004:fitting_power-law} a similar experiment is
        reported. For a fit of the logarithmically binned empirical
        probability distribution the authors find a systematical bias of
        29\,\%. I cannot reconstruct such a strong bias, instead I get a bias
        of 5\,\% only. Probably the quality of this method depends on the
        details of the binning procedure.} & 2.397 & 0.080 \\
      fit on cumulative empirical distribution with logarithmic binning & 2.544 & 0.127 \\
      maximum likelihood & 2.500 & 0.016 \\
      \hline\hline
    \end{tabular}    
  \end{minipage}
\end{table}

There are two ways to deal with the sparseness in the tail of the empirical
distribution, logarithmic binning and considering the empirical cumulative
distribution $\hat P(k)$ instead of $\hat p(k)$. The cumulative probability
distribution of (\ref{eq:p1}) is defined by
\begin{equation}
  P(k) = \sum_{i=k}^\infty  \frac{i^{-\gamma}}{\zeta(\gamma, 1)} \,.
\end{equation}
If $p(k)$ has a power-law tail with exponent $\gamma$ then $P(k)$ follows
approximately a power-law with exponent $\gamma-1$ because for $k\gg1$ the
distribution $P(k)$ can be approximated by
\begin{equation}
  P(k) \approx \int_{k}^\infty  \frac{i^{-\gamma}}{\zeta(\gamma, 1)} \,\D i= 
  \frac{k^{1-\gamma}}{(\gamma -1)\zeta(\gamma, 1)}\,.
\end{equation}
The empirical cumulative probability distribution is given by
\begin{equation}
  \hat P(k) = \frac{1}{N} \sum_{i=1}^N \Id{m_i\ge k}\,.
\end{equation}
It is less sensitive to the noise in the tail of the distribution and
therefore a fit of
\begin{equation}
  (x_i, y_i) = (\ln k, \ln\hat P(k))
  \quad\text{for all $k\in\N$ with $\hat P(k)>0$}
\end{equation}
to a straight line gives much better estimates for the exponent, see
\figurename~\ref{fig:graphical}\,b. But there is still a small bias to too
small values and the distribution of this estimate is rather broad, see
\tablename~\ref{tab:estimate_statistics}.

\sloppy Logarithmic binning reduces the noise in the tail of the empirical
distributions $\hat p(k)$ and $\hat P(k)$ by merging data points into
groups. By introducing the logarithmically scaled boundaries
\begin{equation}
  b_i = \round c^i\quad\text{with some $c>1$}
\end{equation}
\fussy (The function $\round x$ rounds $x$ to the nearest integer.) a linear
least squares fit is performed to
\begin{align}
  (x_i, y_i) & = \left(
    \ln \frac{b_i+b_{i+1}-1}{2} , 
    \ln \sum_{k=b_i}^{b_{i+1}-1} \frac{\hat p(k)}{b_{i+1}-b_i}
  \right)\\
  \intertext{or}
  (x_i, y_i) & = \left(
    \ln \frac{b_i+b_{i+1}-1}{2} , 
    \ln \sum_{k=b_i}^{b_{i+1}-1} \frac{\hat P(k)}{b_{i+1}-b_i}
  \right)\,,
\end{align}
respectively. As a consequence of the binning the width of the distribution of
the estimate $\hat\gamma$ of the power-law exponent $\gamma$ is reduced, see
\figurename~\ref{fig:graphical}\,c, \ref{fig:graphical}\,d and
\tablename~\ref{tab:estimate_statistics}. According to the numerical
experiments a fit of the logarithmically binned cumulative distribution gives
the best results among graphical methods. It shows the smallest systematic
bias.\fussy

All the methods that have been considered so far have a common weakness.  In
the deviation of (\ref{eq:a_0}) and (\ref{eq:a_1}) it was assumed that the
standard deviation of the distribution of the error in $y_i$ is the same for
all data points $(x_i, y_i)$. But this is obviously not the case. For fixed
$k$ the empirical distribution $\hat p(k)$ is a random variable with mean
$p_1(k;\gamma)$ and standard deviation
$\sqrt{p_1(k;\gamma)(1-p_1(k;\gamma))/N}$. For the corresponding data on a
logarithmic scale the standard deviation is approximately given by the
quotient
\begin{equation}
  \label{eq:sigma_log}
  \frac{\sqrt{p_1(k;\gamma)(1-p_1(k;\gamma))/N}}{p_1(k;\gamma)} =
  \sqrt{\frac{1-p_1(k;\gamma)}{Np_1(k;\gamma)}} \,.
\end{equation}
A power-law distribution $p_1(k; \gamma)$ is a monotonically decreasing
function and therefore (\ref{eq:sigma_log}) is an increasing function of
$k$. Because the variation of the statistical error is not taken into account,
the distribution of the estimate $\hat\gamma$ is very broad.

Methods that deal with the cumulative distribution have an additional
weakness. Cumulation has the side-effect that the statistical errors in $y_i$
are not independent any more, which violates another assumption of the
deviation of (\ref{eq:a_0}) and (\ref{eq:a_1}).

To sum up, estimates of exponents of power-law distributions based on a linear
least squares fit are intrinsically inaccurate and lack a sound mathematical
justification.

\section{Maximum likelihood estimators}
\label{sec:ml}

Maximum likelihood estimators offer a solid alternative to graphical methods.
Let $p(k;\theta)$ denote a single parameter probability distribution. The
maximum likelihood estimator $\hat\theta_N$ for the unknown parameter based on
a sample $m_1, m_2,\dots, m_N$ of size $N$ is given by
\begin{equation}
  \label{eq:ml_est}
  \hat\theta_N = \argmax_\theta [L(\theta)] = 
  \argmax_\theta [\ln L(\theta)] \,,
\end{equation}
where
\begin{equation}
  L(\theta) = \prod_{i=1}^N p(m_i; \theta)
\end{equation}
denotes the likelihood function.  In the limit of asymptotically large samples
and under some regularity conditions maximum likelihood estimators share some
desirable features
\cite{Bain:Engelhardt:2000:Statistics,Pawitan:2001:likelihood}.
\begin{itemize}
\item The estimator $\hat\theta_N$ exists and is unique.
\item The estimator $\hat\theta_N$ is consistent, that means for every
  $\varepsilon>0$ 
  \begin{equation}
    \lim_{N\to\infty} \Prob{|\hat\theta_N - \theta|<\varepsilon} = 1\,,
  \end{equation}
  where $\Prob{|\hat\theta_N - \theta|<\varepsilon}$ denotes the probability
  that the difference \mbox{$|\hat\theta_N - \theta|$} is less than
  $\varepsilon$.
\item The estimator $\hat\theta_N$ is asymptotically normal with mean $\theta$
  and variance
  \begin{equation}
    \label{eq:var}
    (\Delta \hat\theta_N)^2 \doteq 
    \left(
      N\, \E{\left(\frac{\D}{\D\theta} \ln p(k;\theta)\right)^2}
    \right)^{-1}\,,
  \end{equation}
  where $\E{\cdot}$ indicates the expectation value of the quantity in the
  brackets.
\item Maximum likelihood estimators have asymptotically minimal variance among
  all asymptotically unbiased estimators. One says, they are asymptotically
  efficient.
\end{itemize}

\section{Maximum likelihood estimators for genuine power-laws}
\label{sec:pure}

The most general discrete genuine power-law distribution has a lower as well
as an upper bound and is given by
\begin{equation}
  \label{eq:p_k_min_max}
  p_{k_\mathrm{min}, k_\mathrm{max}}(k; \gamma) = 
  \frac{k^{-\gamma}}{\zeta(\gamma, k_\mathrm{min}, k_\mathrm{max})}
\end{equation}
for $k\in\N$ with $k_\mathrm{min}\le k < k_\mathrm{max}$. Where the
non-standard notation
\begin{equation}
  \zeta(\gamma, k_\mathrm{min}, k_\mathrm{max}) \definedas
  \zeta(\gamma, k_\mathrm{min}) - \zeta(\gamma, k_\mathrm{max}) 
\end{equation}
has been introduced. If the upper bound is missing the distribution
\begin{equation}
  \label{eq:p_k_min}
  p_{k_\mathrm{min}}(k; \gamma) = 
  \frac{k^{-\gamma}}{\zeta(\gamma, k_\mathrm{min})}
\end{equation}
has to be considered for $k\in\N$ with $k\ge k_\mathrm{min}$. The
distributions (\ref{eq:p_k_min_max}) and (\ref{eq:p_k_min}) are
generalizations of (\ref{eq:p1}) and will be useful for the analysis of more
general distributions that show a power-law behavior only in a certain range
but have an arbitrary profile outside the power-law regime. This kind of
distributions will be considered in section~\ref{sec:tail}.

The maximum likelihood estimator $\hat\gamma_N$ for the parameter $\gamma$ of
the distribution (\ref{eq:p_k_min_max}) follows from (\ref{eq:ml_est}) and is
given by
\begin{equation}
  \label{eq:ml_max_min_max}
  \hat\gamma_N = 
  \argmax_\gamma \left[
    -\gamma \left(\sum_{i=1}^N \ln m_i\right) 
    - N\ln \zeta(\gamma, k_\mathrm{min}, k_\mathrm{max})
  \right]
\end{equation}
or equivalently by the implicit equation
\begin{equation}
  \label{eq:ml_zero_min_max}
  \frac{\zeta'(\hat\gamma_N, k_\mathrm{min}, k_\mathrm{max})}
  {\zeta(\hat\gamma_N, k_\mathrm{min}, k_\mathrm{max})}
  + \frac{1}{N}\sum_{i=1}^{N} \ln m_i = 0\,,
\end{equation}
which has to be solved numerically. The prime denotes the derivative with
respect to $\gamma$.  The asymptotic variance of this estimator $\hat\gamma_N$
follows from (\ref{eq:var}) and equals
\begin{multline}
  \label{eq:var_gamma_min_max}
  (\Delta \hat\gamma_N)^2 \doteq   \frac{1}{N} \\
  \null\times\frac{\zeta(\gamma, k_\mathrm{min}, k_\mathrm{max})^2} {\zeta''(\gamma,
    k_\mathrm{min}, k_\mathrm{max}) \zeta(\gamma, k_\mathrm{min},
    k_\mathrm{max})- \zeta'(\gamma, k_\mathrm{min}, k_\mathrm{max})^2} \,.
\end{multline}

\begin{figure}
  \centering
  \includegraphics[width=\linewidth]{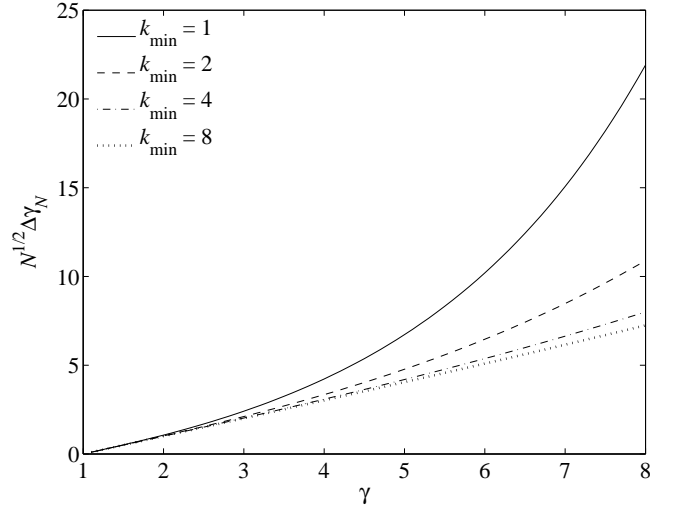}
  \caption{Asymptotic standard deviation for maximum likelihood estimators for
    the exponent of a power-law distribution (\ref{eq:p_k_min}).}
  \label{fig:std_dev}
\end{figure}

\noindent In the limit $k_\mathrm{max}\to\infty$ equations
(\ref{eq:ml_max_min_max}), (\ref{eq:ml_zero_min_max}), and
(\ref{eq:var_gamma_min_max}) give the maximum likelihood estimator and the
asymptotic variance of this estimator for power-law distributions lacking an
upper cut-off (\ref{eq:p_k_min}). A graphical representation of the standard
deviation (\ref{eq:var_gamma_min_max}) in the limit $k_\mathrm{max}\to\infty$
is given in \figurename~\ref{fig:std_dev}. For each fixed $k_\mathrm{min}$ the
quantity $\Delta \hat\gamma_N\sqrt{N}$ grows faster than linear with
$\gamma$. Therefore the larger the exponent $\gamma$ the larger the sample
size that is necessary to get an estimate within a given error bound.

\begin{figure}
  \centering
  \includegraphics[width=\linewidth]{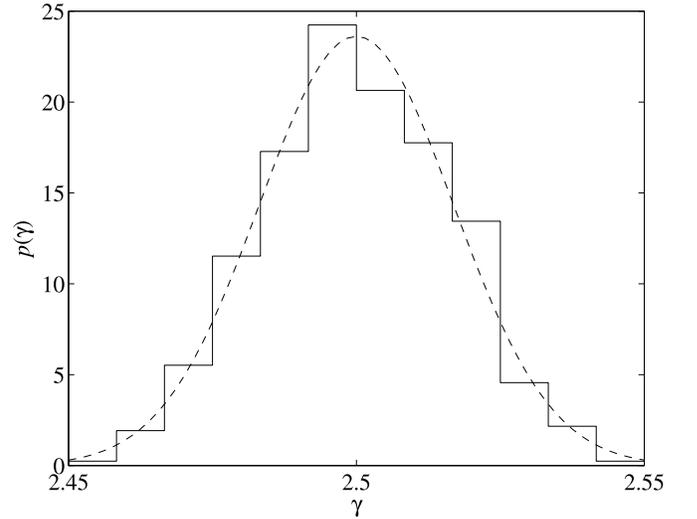}
  \caption{Empirical distribution of the maximum likelihood estimator
    (histogram) versus its theoretical asymptotic distribution, which is given
    by a normal distribution with mean $\gamma=2.5$ and variance
    (\ref{eq:var_gamma_min_max}). The histogram has been obtained from the
    same data as in \figurename~\ref{fig:graphical}.}
  \label{fig:ml}
\end{figure}

If the maximum likelihood method is applied (assuming a distribution
(\ref{eq:p_k_min})) to the same data as in section~\ref{sec:garphical},
numerical experiments show that the estimates for the exponent are much more
precise. The estimate has no identifiable systematic bias, the standard
deviation of the distribution of the estimate is smaller by an order of
magnitude compared to graphical methods, see
\tablename~\ref{tab:estimate_statistics} and \figurename~\ref{fig:ml}.

\section{Maximum likelihood method for general power-law distributions}
\label{sec:tail}

The maximum likelihood procedure outlined in section~\ref{sec:pure} can be
generalized further to distributions $p(k)$ that are no pure power-laws
(\ref{eq:p_k_min_max}) or (\ref{eq:p_k_min}) but follow a power-law within a
certain finite range or follow a power-law in the whole tail of the
distribution and have an arbitrary profile outside the power-law regime.  The
popurse of this section is to establish methods for identifying the power-law
regime and for estimating the exponent of the power-law regime without making
special assumptions about the profile of the probability distribution beyond
the power-law regime.

The main problem for a generalization of the maximum likelihood approach is
that there might be no good hypothesis for the profile of the probability
distribution beyond the power-law regime. To overcome this difficulty the
empirical data set is restricted to a window $k_\mathrm{c\,min}\le m_i
<k_\mathrm{c\,max}$. (The following discussion covers the case of a power-law
tail distributions as well by setting $k_\mathrm{c\,max}=\infty$.) Assuming
that $p(k)$ has a power-law profile for $k_\mathrm{c\,min}\le k
<k_\mathrm{c\,max}$ then the probability distribution of the restricted data
set is given by (\ref{eq:p_k_min_max}) with $k_\mathrm{min}=k_\mathrm{c\,min}$
and $k_\mathrm{max}=k_\mathrm{c\,max}$ (or by (\ref{eq:p_k_min}) with
$k_\mathrm{min}=k_\mathrm{c\,min}$) and some unknown exponent $\gamma$.  This
allows to estimate the power-law exponent by the application of the maximum
likelihood method on the restricted data set of size $N'$ as presented in
section~\ref{sec:pure} \emph{without} making a hypothesis about the profile of
the probability distribution beyond the power-law regime.

In order to apply the maximum likelihood method one has to determine the
cut-off points $k_\mathrm{c\,min}$ and $k_\mathrm{c\,max}$ first. Here it has
to be taken into account that if the window $k_\mathrm{c\,min}\le m_i <
k_\mathrm{c\,max}$ is chosen too large the estimate $\hat\gamma$ is
systematically biased, but on the other hand if it is too small the
statistical error is larger than necessary. In some cases one can make
conservative estimates for $k_\mathrm{c\,min}$ and $k_\mathrm{c\,max}$ by
plotting the empirical probability distribution (\ref{eq:epirical}) on a
double-logarithmic scale. An appropriate window can also be found by
determining estimates $\hat\gamma_{N'}(k_\mathrm{c\,min})$ as a function of
the window and a \mbox{$\chi^2$-test}.

Assuming the empirical data is drawn from a distribution with a power-law tail
(no upper cut-off) the lower cut-off point $k_\mathrm{c\,min}$ can be
determined in the following systematic way. By varying the parameter
$k_\mathrm{c\,min}$ the maximum likelihood approach gives a sequence of
estimates $\hat\gamma_{N'}(k_\mathrm{c\,min})$. If $k_\mathrm{c\,min}$ is very
large the estimate will be quite inaccurate because only a tiny fraction of
the experimental data is taken into account; but the smaller the cut-off
$k_\mathrm{c\,min}$ the more accurate the estimate of the exponent.  If
$k_\mathrm{c\,min}$ approaches the point from above (but is still above) where
the probability distribution starts do differ from a power-law
$\hat\gamma_{N'}(k_\mathrm{c\,min})$ will give a very precise estimate for the
exponent $\gamma$. On the other hand, if $k_\mathrm{c\,min}$ is too small the
hypothesis that the (restricted) empirical data is drawn from a power-law
distribution is violated which causes a significant change of the estimate of
the power-law exponent.

If the empirical data is drawn from a distribution having both a lower
crossover point as well as an upper crossover point a sequence of estimates
$\hat\gamma_{N'}(k_\mathrm{c\,min})$ is determined by restricting the data to
a sliding window $k_\mathrm{c\,min}\le m_i < w
k_\mathrm{c\,min}=k_\mathrm{c\,max}$ with $w>1$. As long as the window lies
completely within the power-law regime the maximum likelihood estimate
obtained from the restricted data set will give a reliable estimate of the
power-law exponent. If the window lies at least partly outside the power-law
regime the estimate is systematically biased.

\JournalPreprint{\begin{figure}[t]}{\begin{figure}[b]}
  \centering\vspace*{11pt}%
  \includegraphics[width=\linewidth]{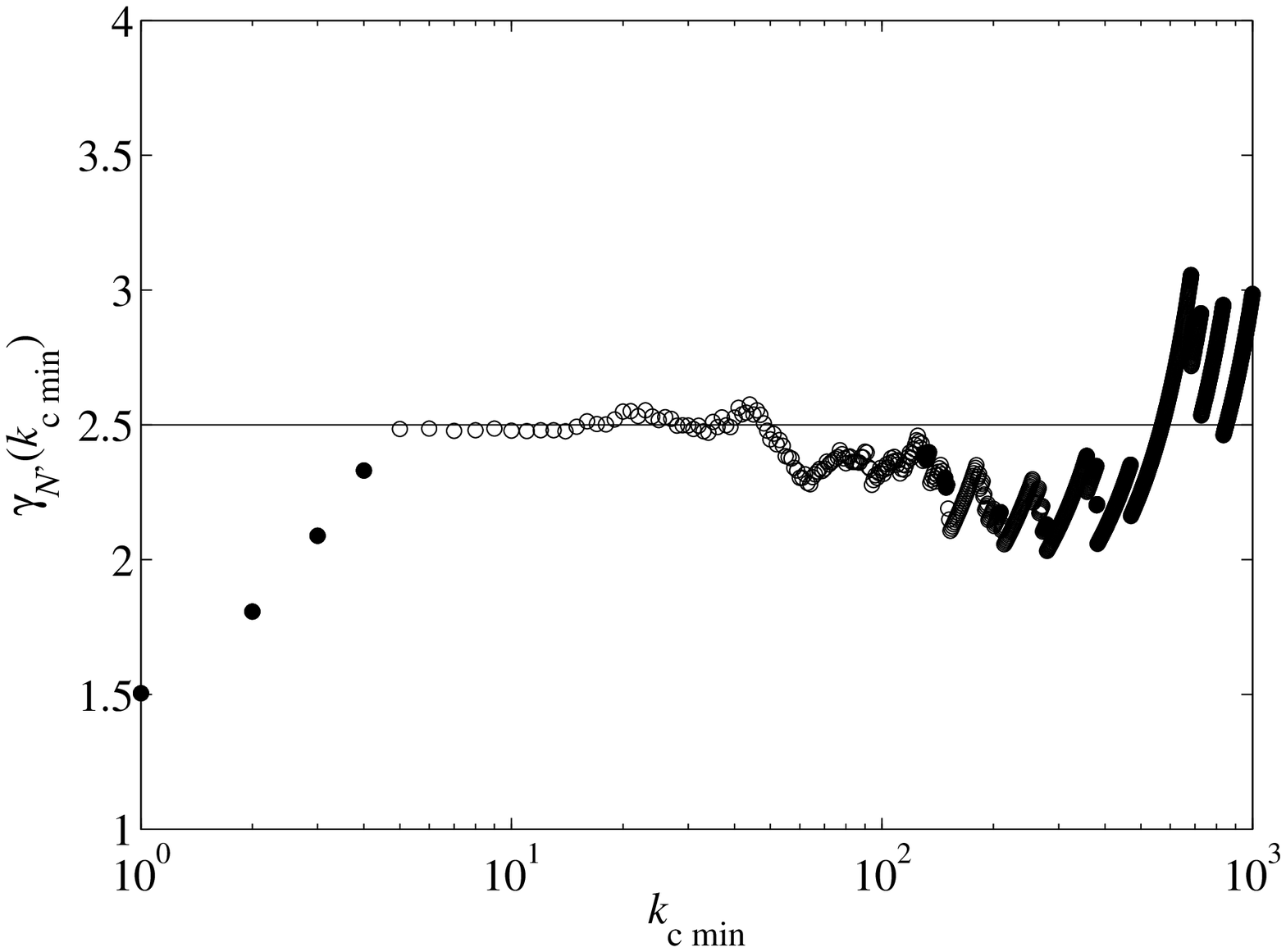}
  \caption{Sequence of estimates $\hat\gamma_{N'}(k_\mathrm{c\,min})$ as a
    function of the cut-off $k_\mathrm{c\,min}$ for a data set of $N=10\,000$
    samples from distribution (\ref{eq:p_cut}). Filled symbols mark where the
    \mbox{$\chi^2$-test} has rejected the hypothesis that the restricted data
    follows a power-law (\ref{eq:p_k_min}) with exponent
    $\gamma=\hat\gamma_{N'}(k_\mathrm{c\,min})$. An error probability of
    $\alpha=0.001$ was chosen.}
  \label{fig:ml_tail}
\end{figure}

To illustrate the procedures outlined above I generated two data sets from two
distributions having a power-law regime. The first data set of $N=10\,000$
samples was drawn from a distribution with a power-law tail which is given by
\begin{equation}
  \label{eq:p_cut}
  p(k) \sim
  \begin{cases}
    5^{-2.5} & \text{for $1\le k\le5$} \\[0.25ex]
    k^{-2.5} & \text{for $k> 5$} \,.
  \end{cases}
\end{equation}
Plotting the sequence of estimates $\hat\gamma_{N'}(k_\mathrm{c\,min})$
against the parameter $k_\mathrm{c\,min}$ reveals the exponent $\gamma=2.5$ as
wells as the crossover point $k=5$ very clearly, see
\figurename~\ref{fig:ml_tail}. The second data set of $N=100\,000$ samples was
drawn from a distribution with two crossover points, viz.
\begin{equation}
  \label{eq:p_cut_2}
   p(k) \sim
  \begin{cases}
    5^{-2.25} & \text{for $1\le k\le 5$} \\[0.25ex]
    k^{-2.25} & \text{for $5\le k\le 100$} \\[0.25ex]
    100^{-2.25}\e^{-0.05(k-100)}& \text{for $k> 100$}\,.
  \end{cases}
\end{equation}
\Figurename~\ref{fig:ml_window} shows the sequence of estimates
$\hat\gamma_{N'}(k_\mathrm{c\,min})$ that had been determined from restricted
data sets of samples within the sliding window $k_\mathrm{c\,min}\le m_i <
5k_\mathrm{c\,min}$. This sequence exhibits a broad plateau that corresponds
to the power-law exponent $\gamma=2.25$. If the window does not lie completely
inside the power-law regime the estimate $\hat\gamma_{N'}(k_\mathrm{c\,min})$
deviates systematically from the known exponent.

\begin{figure}
  \centering
  \includegraphics[width=\linewidth]{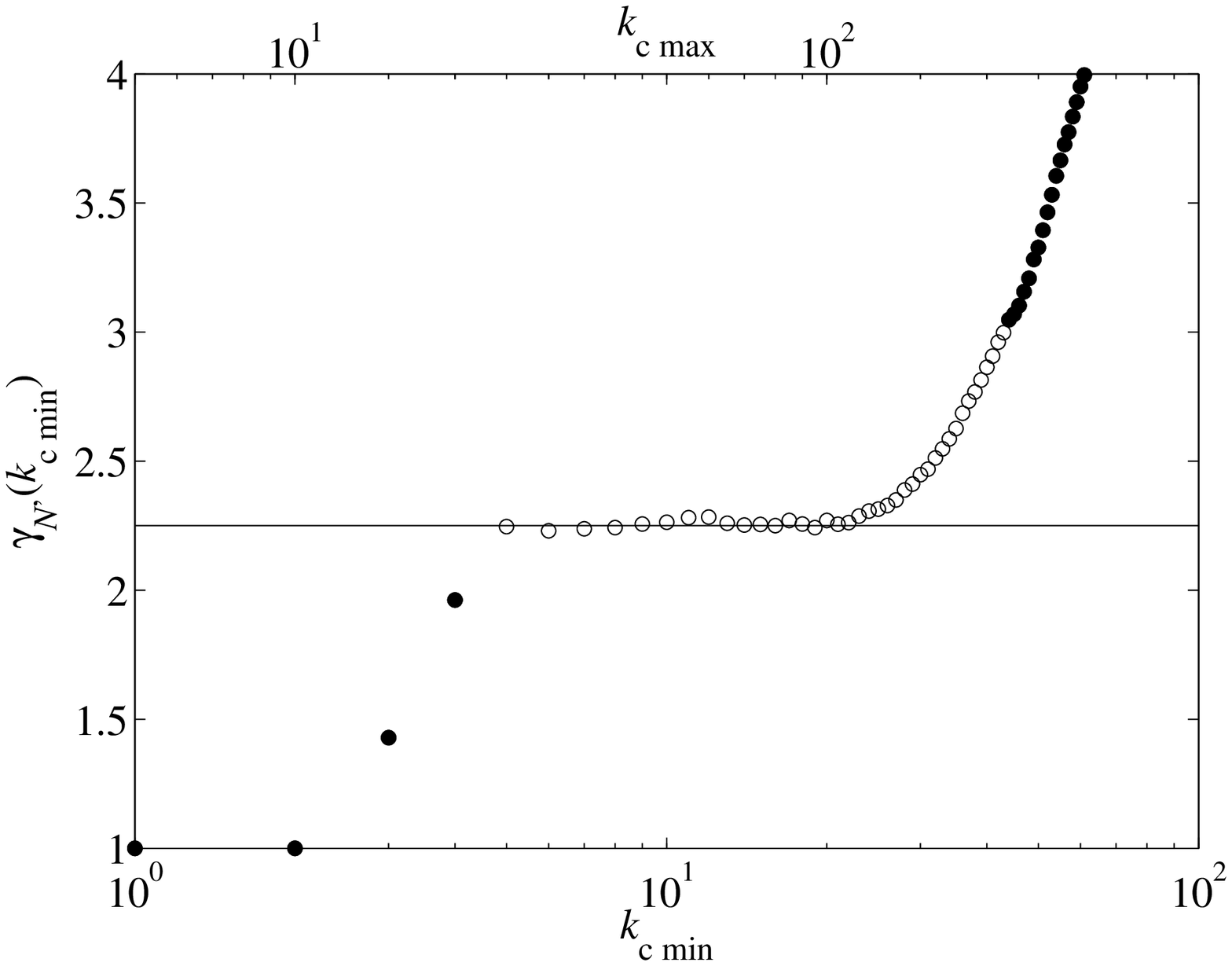}
  \caption{Sequence of estimates $\hat\gamma_{N'}(k_\mathrm{c\,min})$ as a
    function of the lower cut-off $k_\mathrm{c\,min}$ for a data set of
    $N=100\,000$ samples from distribution (\ref{eq:p_cut_2}). Filled symbols
    mark where the \mbox{$\chi^2$-test} has rejected the hypothesis that the
    restricted data follows a power-law (\ref{eq:p_k_min_max}) with exponent
    $\gamma=\hat\gamma_{N'}(k_\mathrm{c\,min})$. An error probability of
    $\alpha=0.001$ and the window width $w=5$ had been chosen.}
  \label{fig:ml_window}
\end{figure}

\sloppy Apart from a visual inspection of the
$\hat\gamma_{N'}(k_\mathrm{c\,min})$ plot the crossover point(s) to the
power-law regime can be determined by means of a \mbox{$\chi^2$-test}. To
apply a \mbox{$\chi^2$-test} the data set has to be divided into some bins and
the following binning turned out to be appropriate: The data is partitioned
into a small number $b$, say $b=6$, of bins. In the case of a distribution
with a power-law tail this means each bin $j$ collects $n_j$ items such that
\def\negsp{\!\!\!\!\!\!\!\!}
\begin{align}
  n_1 & = N' \hat p(k_\mathrm{c\,min}) &
  q_1 & = p_{k_\mathrm{c\,min}}(k_\mathrm{c\,min}; \hat\gamma_{N'}(k_\mathrm{c\,min})) \\
  n_2 & = N' \hat p(k_\mathrm{c\,min}+1) &
  q_2 & = p_{k_\mathrm{c\,min}}(k_\mathrm{c\,min}+1; \hat\gamma_{N'}(k_\mathrm{c\,min})) \\
  & \Vdots &
  & \Vdots \nonumber\\
  \intertext{and finally}
  \label{eq:n_b__q_b}
  n_b & = N' \negsp\sum_{k=k_\mathrm{c\,min}+b-1}^\infty\negsp\hat p(k) &
  q_b & = \negsp \sum_{k=k_\mathrm{c\,min}+b-1}^\infty \negsp p_{k_\mathrm{c\,min}}(k; \hat\gamma_{N'}(k_\mathrm{c\,min})) \,,
\end{align}
where $q_j$ denotes the probability that a data point falls into bin $j$ under
the assumption that the (restricted) data follows the power-law
(\ref{eq:p_k_min}) with $k_\mathrm{min}=k_\mathrm{c\,min}$ and the exponent
$\gamma=\hat\gamma_{N'}(k_\mathrm{c\,min})$. For distributions with a finite
power-law regime the binning procedure can be carried out in a similar way. In
this case the summation index in (\ref{eq:n_b__q_b}) is bounded by
$k_\mathrm{c\,min}+b-1\le k< k_\mathrm{c\,max}$ and the probability
$p_{k_\mathrm{c\,min}, k_\mathrm{c\,max}}(k;
\hat\gamma_{N'}(k_\mathrm{c\,min}))$ has to be considered instead of
$p_{k_\mathrm{c\,min}}(k; \hat\gamma_{N'}(k_\mathrm{c\,min}))$.

The test statistic of the \mbox{$\chi^2$-test} is given by 
\begin{equation}
  \label{eq:chi2}
  c^2  = \sum_{j=1}^b \frac{(n_j - N' q_j)^2}{N' q_j}\,.
\end{equation}
If the to $k_\mathrm{c\,min}\le m_i< k_\mathrm{c\,max}$ restricted data is
given by the power-law (\ref{eq:p_k_min_max}) or (\ref{eq:p_k_min}) with
$k_\mathrm{min}=k_\mathrm{c\,min}$, $k_\mathrm{max}=k_\mathrm{c\,max}$, and
the exponent $\gamma=\hat\gamma_{N'}(k_\mathrm{c\,min})$ then the statistic
$c^2$ follows asymptotically a $\chi^2$-distribution with $\nu=(b-1)$ degrees
of freedom, which is given by
\begin{equation}
  \label{eq:p_chi2}
  p_{\chi^2}(x, \nu) = \frac{x^{\nu/2-1} \e^{-x/2}}{2^{-\nu/2}\,\Upgamma(\nu/2)}\,.
\end{equation}
Let $\chi^2_\alpha$ be the $(1-\alpha)$-quantile of the distribution
(\ref{eq:p_chi2}). The hypothesis that the restricted data is given by the
power-law (\ref{eq:p_k_min_max}) or (\ref{eq:p_k_min}), respectively, with
$k_\mathrm{min}=k_\mathrm{c\,min}$, $k_\mathrm{max}=k_\mathrm{c\,max}$, and
the exponent $\gamma=\hat\gamma_{N'}(k_\mathrm{c\,min})$ is accepted with the
error probability $\alpha$ if $c^2\le\chi^2_\alpha$.  If the window
$k_\mathrm{c\,min}\le m_i < k_\mathrm{c\,max}$ lies not completely within the
power-law regime this hypothesis will be rejected by the \mbox{$\chi^2$-test}
and one can detect the upper crossover point as well as the lower crossover
point (where the power-law loses its validity) in a reliable way, see
\figurename~\ref{fig:ml_tail} and \figurename~\ref{fig:ml_window}.

\section{Computational remarks}

The normalizing factors of the probability distributions
(\ref{eq:p_k_min_max}) and (\ref{eq:p_k_min}) are given by the
Hurwitz-$\zeta$-function. This function is less common than other special
functions and may not be available in the reader's favorite statistical
software package but the GNU Scientific Library \cite{GSL} offers an open
source implementation of this function. A direct calculation of the
Hurwitz-$\zeta$-function by truncating the sum (\ref{eq:H-zeta}) gives
unsatisfactory results.

The maximum likelihood estimator of the exponent can be computed numerically
either by solving (\ref{eq:ml_max_min_max}) or
(\ref{eq:ml_zero_min_max}). Equation (\ref{eq:ml_max_min_max}) has the
advantage that it can be solved without calculating derivatives of the
Hurwitz-$\zeta$-function \cite{Brent:1973:Minimization}, whereas the solution
of (\ref{eq:ml_zero_min_max}) involves its first derivative (e.\,g.\ bisection
method) or even higher derivatives (e.\,g.\ Newton-Raphson method). An
explicit implementation of these derivatives is often not available but may be
calculated numerically.

\section{Conclusion}

Methods based on a least squares fit are not suited to establish estimates for
power-law distribution exponents because least squares fits rely on
assumptions about the data set that are not fulfilled by empirical data from
power-law distributions. In this paper maximum likelihood estimators have been
introduced as a reliable alternative to graphical methods. These estimators
are asymptotically efficient and can be applied to data from a wide class of
distributions having a power-law regime. The crossover points that separate
the power-law regime from the rest of the distribution can be determined by a
procedure based on a \mbox{$\chi^2$-test}.

Finally I would like to mention that the idea to plot a sequence of estimates
$\hat\gamma_{N'}(k_\mathrm{c\,min})$ as shown in \figurename~\ref{fig:ml_tail}
is related to so-called Hill plots
\cite{Hill:1975,Dress:etal:2000:Hill_plot}. The Hill estimator is a maximum
likelihood estimator for the inverse of the exponent of the \emph{continuous}
Pareto distribution
$p(k)=(\gamma-1)(k/k_\mathrm{min})^{-\gamma}/k_\mathrm{min}$, see
\cite{Dress:etal:2000:Hill_plot} for a detailed discussion.

\begin{acknowledgement}
  Work sponsored by the European Community's FP6 Information Society
  Technologies programme under contract IST-001935, EVERGROW. 
\end{acknowledgement}

\bibliographystyle{epj}
\bibliography{estimating_powerlaws}

\end{document}